\documentclass[twocolumn,showpacs,preprintnumbers,amsmath,amssymb,superscriptaddress]{revtex4}

\usepackage{graphicx}
\usepackage{bm}
\usepackage{epsfig}

\allowdisplaybreaks

\begin{document}


\preprint{SFB/CPP-12-04, TTP12-03}
\title{Gauge Coupling Beta Functions in the Standard Model to Three Loops}
\author{Luminita N. Mihaila, Jens Salomon, Matthias Steinhauser}
\affiliation{Institut f\"ur Theoretische Teilchenphysik, Karlsruhe
  Institute of Technology (KIT), D-76128 Karlsruhe, Germany}
%
\begin{abstract}
  In this Letter we compute the three-loop corrections to the beta
  functions of the three gauge couplings
  in the Standard Model of particle physics 
  using the minimal subtraction scheme and
  taking into account Yukawa and Higgs self-couplings.
\end{abstract}


\maketitle


Renormalization group functions are fundamental quantities of each
quantum field theory and play an important role in various
aspects. Besides controlling the energy dependence of parameters and
fields they are also crucial for the resummation of large
logarithms. Furthermore, renormalization group functions are important
for the development of grand unified theories and the
extrapolation of low-energy precision data to high energies, not
accessible by collider experiments.

As far as the strong interaction part of the Standard Model is
concerned the corresponding gauge coupling beta function 
is known up to four-loop 
order~\cite{Gross:1973id,Politzer:1973fx,Jones:1974mm,Caswell:1974gg,Tarasov:1976ef,Egorian:1978zx,Tarasov:1980au,Larin:1993tp,vanRitbergen:1997va,Czakon:2004bu}. 
At three-loop level also the corrections involving two strong and one top quark 
Yukawa coupling have been computed~\cite{Steinhauser:1998cm}.
On the other hand, for the $SU(2)_L \times U(1)_Y$ part only
one-~\cite{Gross:1973id,Politzer:1973fx,Gross:1973ju} and 
two-loop~\cite{Jones:1981we,Fischler:1981is,Machacek:1983tz,Jack:1984vj} 
corrections are available since about 30 years.
(Two-loop corrections are also known
for the Yukawa~\cite{Machacek:1983fi,Fischler:1982du,Jack:1984vj}
and Higgs boson self-coupling~\cite{Machacek:1984zw,Jack:1984vj,Ford:1992pn}
beta function, see also Ref.~\cite{Luo:2002ey}.)
For a general theory based on a simple gauge group the three-loop
corrections to the gauge coupling
beta function have been calculated in Ref.~\cite{Pickering:2001aq}.
In this Letter we provide results for the three-loop gauge coupling
beta functions taking into account all sectors of the Standard Model,
i.e., the gauge, Yukawa and Higgs boson self-couplings.

Let us in a first step define the beta functions. We denote the three
gauge couplings by $\alpha_1$, $\alpha_2$ and $\alpha_3$ and adopt a
$SU(5)$-like normalization with
\begin{eqnarray}
  \alpha_1 &=& \frac{5}{3}\frac{\alpha}{\cos^2\theta_W}\,,\nonumber\\
  \alpha_2 &=& \frac{\alpha}{\sin^2\theta_W}\,,\nonumber\\
  \alpha_3 &=& \alpha_s\,,
\end{eqnarray}
where $\alpha$ is the fine structure constant, $\theta_W$ the weak
mixing angle and $\alpha_s$ the strong coupling.
In our calculation we consider in addition to the gauge couplings also the
third-generation Yukawa couplings~\footnote{A generalization of the
  final result to non-vanishing Yukawa couplings for the first and
  second generation is straightforward.}  
$\alpha_4=\alpha_t$, $\alpha_5=\alpha_b$ and $\alpha_6=\alpha_\tau$, 
and the Higgs boson self-coupling
$\alpha_7=\lambda/(4\pi)$. $\alpha_x = \alpha m_x^2/(2\sin^2\theta_W M_W^2)$
($x=t, b, \tau$) where $m_x$ and $M_W$ are the fermion and $W$ boson mass,
respectively, and $-\lambda (\Phi^\dagger\Phi)^2$
is the part of the Lagrange density describing the
quartic Higgs self interaction.

The functions $\beta_i$ are obtained from the renormalization constants
of the corresponding gauge couplings that are defined as
$g_i^{\rm bare} = \mu^\epsilon Z_{g_i} g_i$ where $\alpha_i = g_i^2/(4\pi)$. 
Exploiting the fact that the $g_i^{\rm bare}$ are $\mu$-independent
and taking into account that $Z_{g_i}$ may depend on all seven couplings
leads to the following formula
\begin{align}
  \beta_i = - \left[ \epsilon \frac{\alpha_i}{\pi}
    + 2 \frac{\alpha_i}{Z_{g_i}}\sum_{j\not=i}
    \frac{\partial Z_{g_i}}{\partial \alpha_j} \beta_j
  \right] 
  \left( 1 + 2 \frac{\alpha_i}{Z_{g_i}} \frac{\partial Z_{g_i}}{\partial
      \alpha_i} \right)^{-1}\,,
  \nonumber\\
  \label{eq::beta_Z}
\end{align}
where $\epsilon=(4-d)/2$ is the regulator of Dimensional Regularization
with $d$ being the space-time dimension used for the evaluation of
the momentum integrals and the dependence of $\alpha_i$ on the
renormalization scale $\mu$ is suppressed.
From Eq.~(\ref{eq::beta_Z}) it is clear that the renormalization
constants $Z_{g_i}$ ($i=1,2,3$)
have to be computed up to three-loop order.

In the modified minimal subtraction ($\overline{\rm MS}$) renormalization
scheme the perturbative expansion of the gauge coupling
beta functions can be written as
\begin{eqnarray}
  \lefteqn{
    \mu^2 \frac{ {\rm d} }{ {\rm d} \mu^2 } \frac{\alpha_i}{\pi}
    =
    \beta_i(\{\alpha_j\},\epsilon)  = -\epsilon \frac{\alpha_i}{\pi} }
  \nonumber\\&&\mbox{}
  -\left( \frac{\alpha_i}{\pi} \right)^2\left[
    a_i
    + \sum_{j=1}^7 \frac{\alpha_j}{\pi} b_{ij}
    + \sum_{j,k=1}^7 \frac{\alpha_j}{\pi}\frac{\alpha_k}{\pi} c_{ijk}
    + \ldots
    \right]\,.
  \nonumber\\
  \label{eq::beta_def}
\end{eqnarray}
In this Letter we evaluate the three-loop terms
(coefficients $c_{ijk}$) only for the gauge couplings
(i.e. $i=1,2,3$). 
For our calculation
the beta functions for the Yukawa couplings are needed
to the one-loop order 
and the tree-level expression [first term in Eq.~(\ref{eq::beta_def})] is
sufficient for $\beta_\lambda$. 

In the $\overline{\rm MS}$ scheme the beta functions are mass independent
that allows us to use the Standard Model in the unbroken phase as a
framework for our calculation. By construction 
less vertices have to be considered than after spontaneous
symmetry breaking. The electroweak gauge bosons are denoted by
$W$ and $B$ corresponding to the $SU(2)_L \times U(1)_Y$ gauge groups,
respectively.

\begin{figure}[tb]
  \begin{center}
    \begin{tabular}{ccc}
      \includegraphics[width=7em]{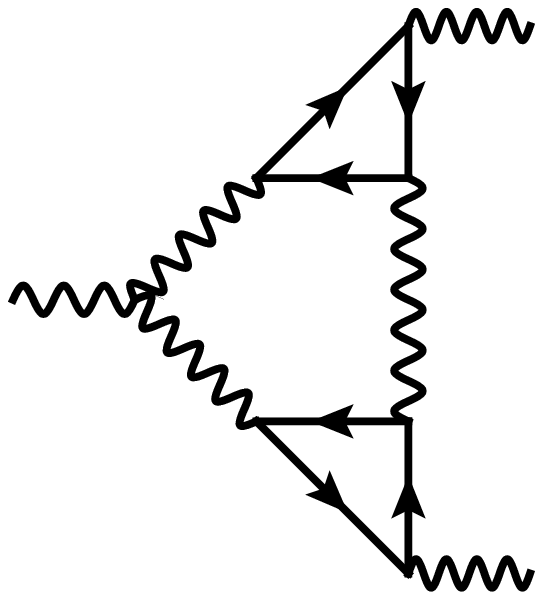} &
      \includegraphics[width=7em]{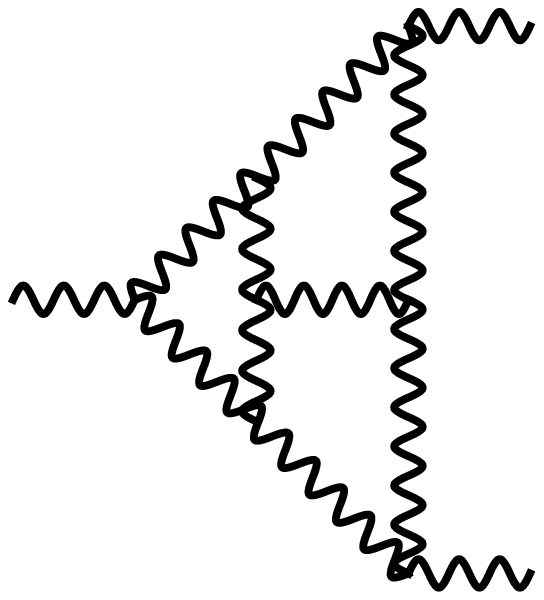} &
      \includegraphics[width=7em]{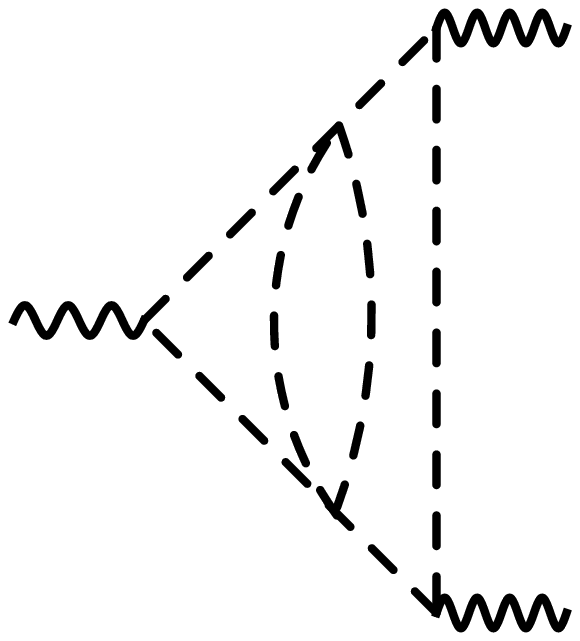} \\
      \\
      (a) & (b) & (c)\\
      \includegraphics[width=7em]{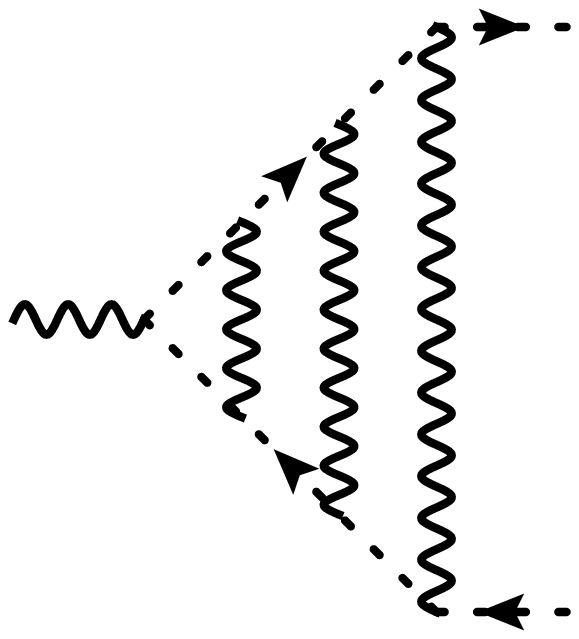} &
      \includegraphics[width=7em]{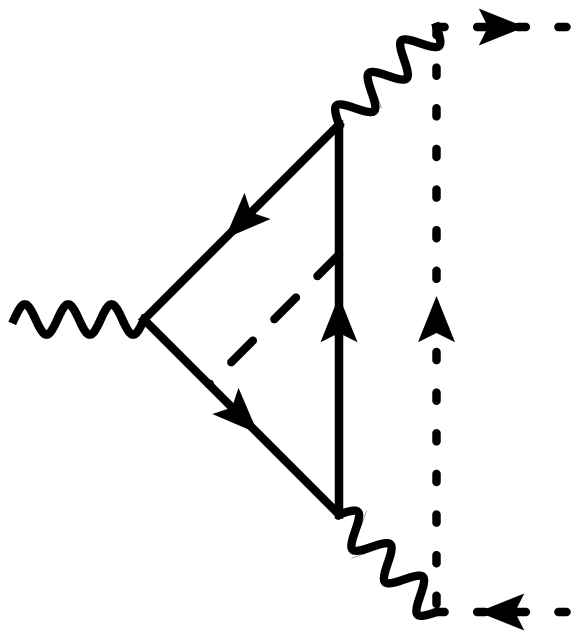}
      \\
      (d) & (e)
    \end{tabular}
    \caption{\label{fig::diags}Sample Feynman diagrams contributing to the
      $W_1W_2W_3$ vertex [(a), (b) and (c)] and the ghost-$W_3$
      vertex [(d) and (e)].
      Dashed, dotted, solid and wavy lines correspond to 
      Higgs boson, ghost, fermion and $W$ boson lines, 
      respectively. Sample diagrams for gluonic Green's functions are
      obtained for (a), (b), (d) and (e) by replacing the $W$ lines
      by gluon lines.}
  \end{center}
\end{figure}

In principle each vertex containing the gauge coupling $g_i$ at tree level 
can be used in order to obtain $Z_{g_i}$ via
\begin{eqnarray}
  Z_{g_i} &=& \frac{ Z_{\rm vert} }{\Pi_{k} \sqrt{Z_{k,{\rm wf}}} }
  \,,
\end{eqnarray}
where $Z_{\rm vert}$ stands for the renormalization constant of the
vertex and $Z_{k,{\rm wf}}$ for the wave function renormalization
constant; $k$ runs over all external particles.
We have computed $Z_{g_3}$ using both the 
ghost-gluon and the three-gluon vertex and $Z_{g_2}$ has been
evaluated with the help of the ghost-$W_3$, the $W_1 W_2 W_3$ and the 
$\phi^+\phi^- W_3$ vertex where
$\phi^\pm$ is the charged component of the Higgs doublet corresponding to the 
Goldstone boson in the broken phase and 
$W_1$, $W_2$ and $W_3$ are the components of the $W$ boson.
Both for $Z_{g_2}$ and $Z_{g_3}$ the different ways lead to the same result
that constitutes a strong check for the correctness of the final
result. 
$Z_{g_1}$ is obtained from vertices containing the $B$ boson. Because of
the Ward identity there is a cancellation of $Z_{\rm vert}$ and 
the factors $\sqrt{Z_{k,{\rm wf}}}$ other than the one corresponding to the $B$
boson. Thus $Z_{g_1}$ is solely computed from 
the wave function renormalization constant of the $B$ boson.
Several three-loop sample diagrams contributing to the considered three-point
functions are shown in Fig.~\ref{fig::diags}. 
Because of the fact that the $\beta_i$ do not depend on any kinematical
scale we evaluate the vertex functions in the limit where one
external momentum is set to zero. In this way all loop-integrals are
mapped to massless two-point functions that up to three loops can be
computed with the help of {\tt MINCER}~\cite{Larin:1991fz}.

An important issue in the present 
calculation is the treatment of $\gamma_5$ within dimensional regularization. 
Non-trivial contributions may arise if in the course of the
calculation two fermion traces occur where both of them contain an
odd number of $\gamma_5$ matrices and four or more $\gamma$
matrices. It is straightforward to see that the Green's functions that
we have chosen for calculating the beta functions contain at most
one-loop triangle sub-diagrams  (see, e.g., Fig.~\ref{fig::diags}(a)).
This could potentially lead to
contributions where a careful treatment of $\gamma_5$ is required.
In our case, however, all these contributions vanish identically due
to anomaly cancellations [see, e.g., Ref.~\cite{Peskin:1995ev}] since
we always sum over all fermions of one generation.
This has also been checked by an explicit calculation~\cite{MSS12}.

Our calculation is based on a high level of automation in order to
avoid errors due to manual interaction.
As a core of our set-up we use the well-tested chain of programs 
that work hand-in-hand: 
{\tt QGRAF}~\cite{Nogueira:1991ex} generates all contributing Feynman
diagrams. The output is passed via
{\tt q2e} to {\tt exp}~\cite{Harlander:1997zb,Seidensticker:1999bb}
that generates the {\tt FORM}~\cite{Vermaseren:2000nd} code. The
latter is processed by {\tt MINCER}~\cite{Larin:1991fz} that
computes the Feynman integrals and outputs
the $\epsilon$ expansion of the result.
A serious problem that had to be overcome in the course of
the present calculation is the enormous number of
diagrams that contribute to the individual renormalization factors.
For example, in the case of the $W_1W_2W_3$ vertex one has about 380\,000
diagrams for the vertex itself and 60\,000 for the $W_3$ self energy
with similar numbers for the other Green's functions. Thus, in total
more than one million diagrams have to be computed.
In order to handle such an amount of diagrams we have decided to
automatically split the original problems into blocks containing
only of the order of 1000 Feynman diagrams. 
In total we end up with a CPU time of about 100 days on a single
core. Since our calculation is highly parallelizable, the final 
wall-clock time is about one day on 100 cores.

Note that at three-loop
level all sectors of the Standard Model contribute to the $\beta$
functions of the gauge couplings.
Thus the huge number of diagrams mainly results from the numerous
interaction vertices in the Standard Model. 
In the set-up described above the strong interaction part of the
Standard Model has already been used for a variety
of calculations. We have added the electroweak part
by establishing an interface to a model file of
{\tt FeynArts}~\cite{Hahn:2000kx}. It generates all Standard Model
Feynman rules in a format readable by {\tt QGRAF} and {\tt q2e}.
The model file for {\tt FeynArts} has been generated using 
{\tt FeynRules}~\cite{Christensen:2008py}.

In the following we present results for $\beta_1$, $\beta_2$ and
$\beta_3$ up to three loops. In order to keep the expressions
compact we set the Yukawa couplings $\alpha_b$ and $\alpha_\tau$ to
zero. Also, we use unit Cabibbo-Kobayashi-Maskawa matrix in our calculations.
The complete result with $\alpha_b$ and $\alpha_\tau$ kept non-zero can be found
elsewhere~\cite{MSS12}. Omitting the $\mu$ dependence in the arguments 
of $\alpha_i$ (and taking the limit $\epsilon\to 0$) we obtain
\begin{widetext}
\begin{align}
  \beta_1 &= 
  \left(\frac{\alpha_1}{\pi}\right)^2 \Bigg\{
    \frac{1}{40} + \frac{n_G}{3} 
    + \frac{\alpha_1}{\pi} \left(\frac{9}{800} + \frac{19n_G}{240}\right)
    + \frac{\alpha_2}{\pi} \left(\frac{9}{160} + \frac{3n_G}{80}\right)
    + \frac{\alpha_3}{\pi} \frac{11n_G}{60}
    + \left(\frac{\alpha_1}{\pi}\right)^2 
    \left(  \frac{489}{512000} -\frac{29 n_G}{2400} - \frac{209 n_G^2}{8640} \right)
    \nonumber\\&\mbox{}
    + \frac{\alpha_1}{\pi}\frac{\alpha_2}{\pi} 
    \left(  \frac{783}{51200} - \frac{7 n_G}{6400} \right)
    - \frac{\alpha_1}{\pi}\frac{\alpha_3}{\pi} \frac{137n_G}{14400}
    + \left(\frac{\alpha_2}{\pi}\right)^2 
    \left( \frac{3401}{20480} + \frac{83n_G}{1920} - \frac{11n_G^2}{960}  \right) 
    - \frac{\alpha_2}{\pi}\frac{\alpha_3}{\pi} \frac{n_G}{320}
    \nonumber\\&\mbox{}
    + \left(\frac{\alpha_3}{\pi}\right)^2
    \left(\frac{275n_G}{576} - \frac{121n_G^2}{2160}\right)
    + n_t \frac{\alpha_t}{\pi} \left[
      - \frac{17}{160}
      - \frac{\alpha_1}{\pi} \frac{2827}{51200}
      - \frac{\alpha_2}{\pi} \frac{471}{2048}
      - \frac{\alpha_3}{\pi} \frac{29}{320}
      + \frac{\alpha_t}{\pi}
      \left( \frac{339}{5120} + \frac{303 n_t}{2560}\right)
    \right]
    \nonumber\\&\mbox{}
    + \frac{\lambda}{4\pi^2} \left(
      \frac{\alpha_1}{\pi}   \frac{27}{3200} 
      + \frac{\alpha_2}{\pi} \frac{9}{640} 
      - \frac{\lambda}{4\pi^2} \frac{9}{320}
      \right)
    \Bigg\}\,,
  \nonumber\\
  \beta_2 &= 
  \left(\frac{\alpha_2}{\pi}\right)^2 \Bigg\{
    -\frac{43}{24} + \frac{n_G}{3} 
    + \frac{\alpha_1}{\pi} \left(\frac{3}{160} + \frac{n_G}{80}\right)
    + \frac{\alpha_2}{\pi} \left(-\frac{259}{96} + \frac{49n_G}{48}\right)
    + \frac{\alpha_3}{\pi} \frac{n_G}{4}
    + \left(\frac{\alpha_1}{\pi}\right)^2 
    \left(  \frac{163}{102400} -\frac{7 n_G}{960} - \frac{11 n_G^2}{2880} \right)
    \nonumber\\&\mbox{}
    + \frac{\alpha_1}{\pi}\frac{\alpha_2}{\pi} 
    \left(  \frac{561}{10240} + \frac{13 n_G}{1280} \right)
    - \frac{\alpha_1}{\pi}\frac{\alpha_3}{\pi} \frac{n_G}{960}
    + \left(\frac{\alpha_2}{\pi}\right)^2 
    \left(  -\frac{667111}{110592} + \frac{1603 n_G}{432} - \frac{415
      n_G^2}{1728} \right) 
    + \frac{\alpha_2}{\pi}\frac{\alpha_3}{\pi} \frac{13 n_G}{64}
    \nonumber\\&\mbox{}
    + \left(\frac{\alpha_3}{\pi}\right)^2
    \left(\frac{125n_G}{192} - \frac{11n_G^2}{144}\right)
    + n_t \frac{\alpha_t}{\pi} \left[
      - \frac{3}{32}
      - \frac{\alpha_1}{\pi} \frac{593}{10240}
      - \frac{\alpha_2}{\pi} \frac{729}{2048}
      - \frac{\alpha_3}{\pi} \frac{7}{64}
      + \frac{\alpha_t}{\pi}
      \left( \frac{57}{1024} + \frac{45 n_t}{512}\right)
    \right]
    \nonumber\\&\mbox{}
    + \frac{\lambda}{4\pi^2} \left(
      \frac{\alpha_1}{\pi}   \frac{3}{640} 
      + \frac{\alpha_2}{\pi} \frac{3}{128} 
      - \frac{\lambda}{4\pi^2} \frac{3}{64}
      \right)
    \Bigg\}\,,
  \nonumber\\
  \beta_3 &=
  \left(\frac{\alpha_3}{\pi}\right)^2 \Bigg\{
    -\frac{11}{4} + \frac{n_G}{3} 
    + \frac{\alpha_1}{\pi} \frac{11n_G}{480}
    + \frac{\alpha_2}{\pi} \frac{3n_G}{32}
    + \frac{\alpha_3}{\pi} \left(-\frac{51}{8} + \frac{19n_G}{12}\right)
    + \left(\frac{\alpha_1}{\pi}\right)^2 
    \left(  -\frac{13 n_G}{7680} - \frac{121 n_G^2}{17280}  \right)
    \nonumber\\&\mbox{}
    - \frac{\alpha_1}{\pi}\frac{\alpha_2}{\pi} \frac{n_G}{2560}
    + \frac{\alpha_1}{\pi}\frac{\alpha_3}{\pi} \frac{77 n_G}{2880}
    + \left(\frac{\alpha_2}{\pi}\right)^2 
    \left(  \frac{241 n_G}{1536} - \frac{11 n_G^2}{384}  \right) 
    + \frac{\alpha_2}{\pi}\frac{\alpha_3}{\pi} \frac{7 n_G}{64}
    \nonumber\\&\mbox{}
    + \left(\frac{\alpha_3}{\pi}\right)^2
    \left(-\frac{2857}{128} + \frac{5033n_G}{576} - \frac{325n_G^2}{864}\right)
    + n_t \frac{\alpha_t}{\pi} \left[
      -  \frac{1}{8}
      - \frac{\alpha_1}{\pi} \frac{101}{2560}
      - \frac{\alpha_2}{\pi} \frac{93}{512}
      - \frac{\alpha_3}{\pi} \frac{5}{8}
      + \frac{\alpha_t}{\pi}
      \left( \frac{9}{128} + \frac{21 n_t}{128}\right)
    \right]
    \Bigg\}\,,
\end{align}
\end{widetext}
where $n_G$ labels the number of generations and $n_t$ the number of heavy
up-type quarks. In the Standard Model we have
$n_G=3$ and $n_t=1$. The result including a fourth generation~\footnote{The
  extension to more generations is obvious. Of course, for phenomenologically
  interesting mass parameters of a fourth quark generation all
  Yukawa couplings should be kept non-zero~\cite{MSS12}.} 
(with Yukawa coupling $\alpha_{t^\prime}$)
is obtained with the replacements 
$(\alpha_t n_t)^n \to (\alpha_t + \alpha_{t^\prime})^n$ ($n=1,2$) and
$\alpha_t^2 n_t \to \alpha_t^2 + \alpha_{t^\prime}^2$.
It is interesting to note that, although the two-loop expression is
$\lambda$-independent, the three-loop term of  
$\beta_1$ and $\beta_2$ contains both linear and quadratic terms in $\lambda$.
The latter arise from diagrams as those in Fig.~\ref{fig::diags}(c).

There are several checks on the correctness of our result. The
interface between {\tt FeynArts} and {\tt QGRAF}/{\tt q2e} has been
checked by evaluating several one- and two-loop results that are
known in the literature. We have even considered quantities within the
Minimal Supersymmetric Standard Model, like the relation between the
squark masses within one generation, which is quite involved in case
electroweak interactions are kept non-zero.
We have furthermore
reproduced the two-loop results in the literature both
for the gauge~\cite{Jones:1981we,Fischler:1981is,Machacek:1983tz,Jack:1984vj}
and the Yukawa coupling beta 
functions~\cite{Machacek:1983fi,Fischler:1982du,Jack:1984vj} that constitutes
a strong check 
on the correctness of the Feynman rules and the general set-up.
This is also true for the three-loop corrections to the $BBB$
vertex which we have verified to be zero.
Also the three-loop corrections of order $\alpha_3^3\alpha_t$ to $\beta_3$
from Ref.~\cite{Steinhauser:1998cm} have been reproduced.
All our calculations have been performed for three general
gauge parameters, one for each gauge group. Whereas the
renormalization constants for the vertices and wave functions still
depend on the gauge parameters the $Z_{g_i}$ are independent leading
to gauge parameter independent results for $\beta_i$.
In a further check we have ensured that in the vertex diagrams no
infra-red divergence is introduced although one external momentum is
set to zero~\footnote{The two-point functions are infra-red safe.}.
This is done by assigning a non-vanishing mass $m$ to
internal lines (a common mass for all particles is sufficient) and
performing an asymptotic expansion~\cite{Smirnov:2002pj} in the limit
$q^2\gg m^2$ where $q$ is the non-vanishing external momentum of the
vertex diagram. Asymptotic expansion for this limit is automated in
the program {\tt exp} and thus can be performed with the set-up
described above. It is sufficient to restrict to the leading term in 
$m^2/q^2$ and check that no $\ln(m^2/\mu^2)$ terms appear in the final
result. Nevertheless the calculation becomes significantly more complex;
some diagrams develop up to 35 sub-diagrams when applying the rules of
asymptotic expansion.
With this method we have explicitly checked that the $W_1W_2W_3$ and
three-gluon vertex
are free from infra-red divergences. Since the
results agree with the ones obtained from the other vertices also
the latter are infra-red safe.

In order to estimate the numerical effect of the new terms we use
the experimentally measured values for $\alpha_i(M_Z^2)$ ($i=1,2,3$)
and run to a higher scale $\mu$ using two or three loops for the gauge 
and two loops for top Yukawa beta functions.
For $\mu=2$~TeV the relative difference between two and three loops amounts
to 0.003\%, 0.010\% and 0.005\% for $\alpha_1$, $\alpha_2$ and $\alpha_3$,
respectively. In the case of $\alpha_3$ the shift is significantly smaller
than the one introduced by the experimental uncertainty at the $Z$ boson mass
scale. On the other hand, for $\alpha_1$ and $\alpha_2$ the three-loop effect
is of the same order of magnitude compared to the experimental uncertainty.
Similar conclusions hold for $\mu=10^{16}$~GeV where a relative shift between
two- and three-loop running amounts 
to 0.012\%, 0.027\% and 0.004\% for $\alpha_1$, $\alpha_2$ and $\alpha_3$,
respectively.

To conclude, in this Letter we have computed the three-loop
corrections to the gauge coupling beta functions in the Standard
Model. Whereas Yukawa corrections are already present at two-loop
order the Higgs boson self-coupling appears for the first time at
three loops.
The numerical effect of the new terms is small, however, in the case of
$\alpha_1$ and $\alpha_2$ it is comparable to the experimental uncertainties.

The method used for the current calculation can also be applied to the
calculation of the beta functions of the Yukawa and Higgs boson
self-coupling. However, in the case of the Yukawa couplings the issue of
$\gamma_5$ is likely to be more serious. For the Higgs boson self-coupling, on
the other hand, one has to consider four-point functions and thus a mapping to
massless two-point functions introduces infra-red divergences already at
one-loop order.  A promising method would be to assign a common mass to all
fields and set all external momenta to zero (see, e.g.,
Ref.~\cite{Misiak:1994zw}).



Acknowledgements. 
We are grateful to Konstantin Chetyrkin and Johann
K\"uhn for useful comments, and we thank David Kunz for tests
concerning the interface between {\tt FeynArts} and {\tt q2e}.
This work is supported by DFG through SFB/TR~9
``Computational Particle Physics''.




\end{document}